\documentclass[twocolumn,preprintn/Users/antiasanchezbotana/Box Sync/draft_scn_mgo/PRL_SUBMISSION/ScN_prb_submission_v7.texumbers,floatfix,showpacs,prb,superscriptaddress]{revtex4}
\usepackage{graphicx}
\usepackage{float}
\usepackage{dcolumn} 
\usepackage{bm}
\usepackage{epsfig}
\usepackage{natbib}
\usepackage{amsmath}
\usepackage{mathtools}
\usepackage{listings}
\usepackage{longtable}
\usepackage{csquotes}
\usepackage{gensymb}

\begin{document} 

\title{Nitride Multilayers as a Platform for Parallel \\
    Two-Dimensional Electron-Hole Gases: MgO/ScN(111)}

\author{Antia S. Botana}
\affiliation{Department of Physics, University of California Davis,  Davis, California 95616, USA}
\author{Victor Pardo}
\affiliation{Departamento de F\'isica Aplicada, Universidade de Santiago de Compostela, E-15782 Santiago de Compostela, Spain}
\affiliation{Instituto de Investigaci\'{o}ns Tecnol\'{o}xicas, Universidade de Santiago de Compostela, E-15782 Campus Sur s/n,  Santiago de Compostela, Spain}  
\author{Warren E. Pickett}
\email{wepickett@ucdavis.edu}
\affiliation{Department of Physics, University of California Davis,  Davis, California 95616, USA}
\date{\today}
\begin{abstract}
At interfaces between insulating oxides LaAlO$_3$ and SrTiO$_3$, a two dimensional 
electron gas (2DEG) has been observed and well studied, while the predicted hole gas
(2DHG) has not been realized due to the strong tendency of holes in oxygen $2p$
orbitals to localize.
Here we propose, via ab initio calculations, an unexplored class of materials for the realization of parallel two 
dimensional (2D), two carrier (electron+hole) gases: nitride-oxide heterostructures, 
with (111)-oriented ScN and MgO as the specific example.
Beyond a critical thickness of five ScN layers, this interface hosts spatially separated 
conducting Sc-$3d$ electrons and N-$2p$ holes, each confined to $\sim$two atomic layers
-- the transition metal nitride provides both gases. 
A guiding concept is that the N$^{3-}$ anion should promote robust two carrier 2D hole
conduction compared to that of O$^{2-}$; metal mononitrides are mostly metallic and 
even superconducting while most metal monoxides are insulating. A second positive
factor is that the density of transition metal ions, hence of a resulting 
2DG, is about three times larger for a rocksalt (111) interface than for a
perovskite(001) interface.
Our results, including calculation of Hall coefficient and thermopower for
each conducting layer separately, provide guidance for new exploration, 
both experimental and theoretical, 
on nitride-based conducting gases that should promote study of long sought 
exotic states viz. new excitonic phases and distinct, nanoscale parallel 
superconducting nanolayers. 
\end{abstract}
\maketitle

\section{background}

The unexpected magnetic and electronic phases appearing at the interface between two perovskite oxides have been studied extensively during the last decade.\cite{lao_sto_conducting, mannhart, caviglia, lao_sto_ferromagnetic, lao_sto_superconducting} The discovery of a two dimensional electron gas (2DEG) at the interface between LaAlO$_3$/ SrTiO$_3$(LAO/STO) by Ohtomo and Hwang\cite{lao_sto_conducting} stimulated excitement for the design of similar heterostructures and their possible use in electronic devices.\cite{oxide_electronics_1, oxide_electronics_2}
The further realization that electronic reconstruction could provide a parallel, nanoscale 2DEG and 2D hole gas (2DHG) provided a more exotic vista,\cite{pentcheva_hole_electron, electron_hole_2d} including two band, two carrier magnetic or superconducting systems, as well as long sought excitonic condensates.\cite{millis,el_hole_cond} Substantial theorizing has been waiting favorable experimental platforms for progress.  Unfortunately, at oxide interfaces the hole gas has never materialized. Due to the proclivity of holes to occupy O $2p$ orbitals, the $p$-type interface is always observed to be non-conducting, eliminating the many possibilities provided by two, or many, 2DEGs and 2DHGs separated by only 2-3 nm.\cite{hole_mobility, lvo_sto_p_type_ins} 

The difficulty in obtaining conducting $p$-type interfaces in oxides suggests using related but less electronegative ions, bringing to mind nitrides. Whereas $3d$ transition metal monoxides are mostly strongly correlated (Mott) insulators, \cite{rivadulla_monoxides} the existing mononitrides are usually conducting and even superconducting.\cite{matthias, sc_review_modern_phys} This tendency toward conductivity while forming similar or even identical structures brings a new dimension to the design of heterostructures of insulators having active interfaces.  

(001) layers of transition metal mononitrides (or monoxides) with rocksalt structure are charge neutral, thus being ineffective in providing the 2D gases of interest here. This layer neutrality, and the challenges provided by localized O $2p$ states, invites innovative material design, especially because layer-by-layer growth provides the experimental capability of concerted theory plus experiment design of novel materials with tailored properties at the nanoscale. 

\begin{figure}
\includegraphics[width=0.9\columnwidth,draft=false]{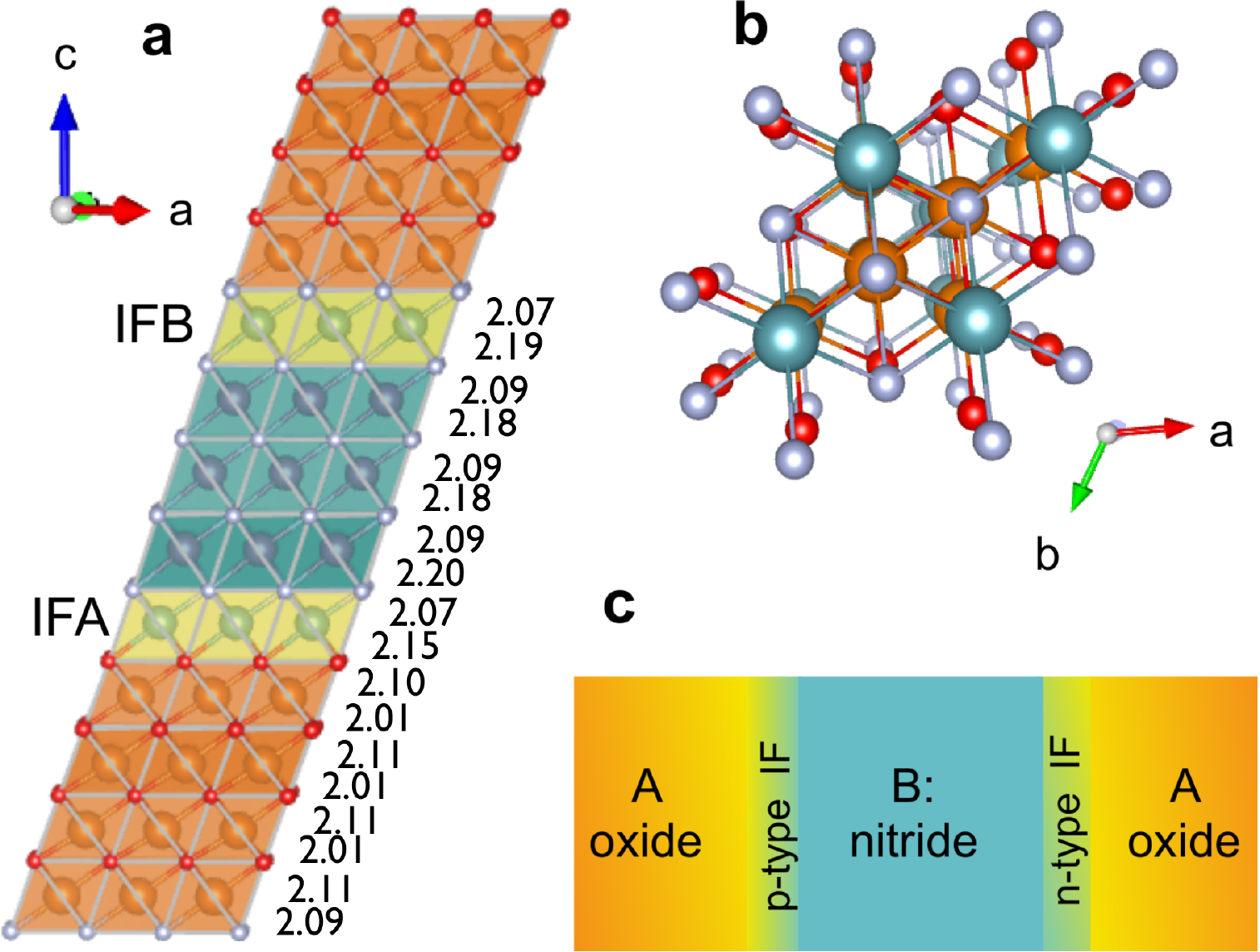}
\caption{(Color online) Structure of ScN/MgO (111) multilayers. (\textbf{a}) Side view of the unit cell of ScN/MgO(111) multilayers five ScN layers thick with Sc atoms in blue, N atoms in gray, Mg atoms in orange, and O atoms in red. Yellow indicates the IF region: IFA, $n$-type (O-Sc) and IFB, $p$-type (N-Mg). The average distances between layers along the \textit{c} axis are shown (in \AA) (\textbf{b}) Triangular-like structure of the multilayers in the $ab$ plane. (\textbf{c}) Sketch of the heterostructure configuration consisting on a thin slab of a transition metal nitride (B, ScN) sandwiched between an oxide (A, MgO)}\label{fig1}
\end{figure}

The simplest polar structure in rocksalt mononitrides is provided by (111) orientation, where oppositely charged metal and anion-atom layers alternately stack along the [111] direction (see Fig. \ref{fig1}). In spite of the widely discussed instability of ideal polar surfaces in a semi-infinite geometry, MgO films with atomically flat (111) surface regions have been reported using layer-by-layer growth on a variety of substrates.\cite{mgo_111, mgo_111_2, mgo_111_3, mgo_111_9, mgo_111_10, mgo_111_11, mgo_111_12, mgo_111_13, hosono}  These atomically flat surfaces are promising to explore the possibility of polar interface engineering using simple metal monoxides and mononitrides with rocksalt structure. 

Here, we explore the use of the narrow gap semiconductor ScN,\cite{gall_2, lambrecht, gw}  sandwiched between highly insulating MgO\cite{mgo_struct,mgo_bulk} spacer layers, as a material to support bilayer electron+hole conducting gases and all the unusual phases that have been anticipated in such heterostructures.  By means of first principles calculations, we have designed and explored MgO/ScN(111) superlattices with varying ScN thickness containing two charge imbalanced interfaces: one $n$-type and one $p$-type. Beyond a ScN thickness threshold of five layers these interfaces host conducting electron+hole gases, both gases lying within but on opposite edges of the ScN layer. Use of the N anion should promote robust two carrier 2D conduction compared to oxides, where holes are prone to localization. ScN is a propitious choice of nitride, since it has been shown that it can be hole-doped in bulk to a similar magnitude of conductivity as natively electron-doped samples.\cite{saha}

\begin{figure}
\includegraphics[width=0.86\columnwidth,draft=false]{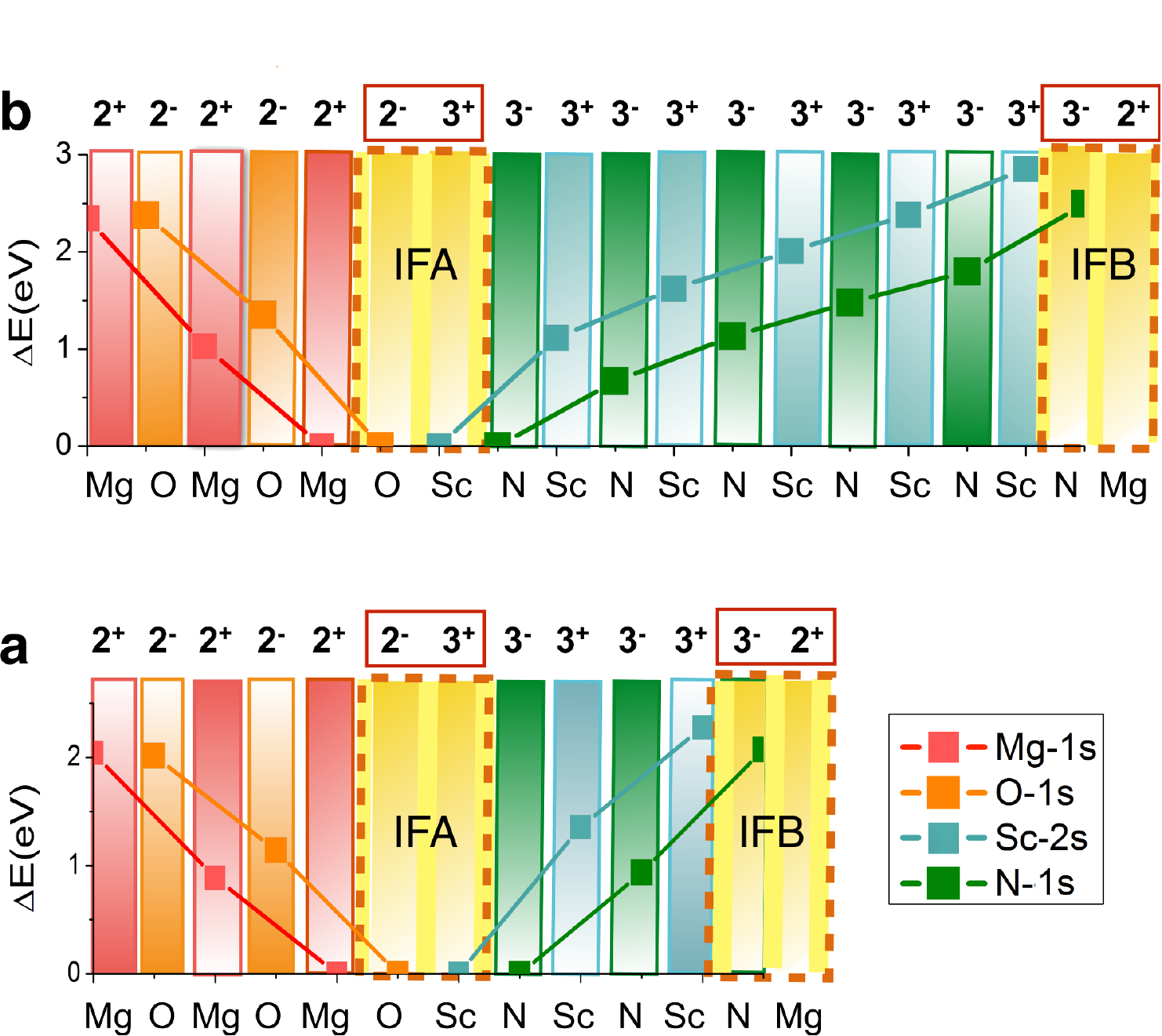}
\caption{(Color online) Potential gradient in the multilayers. Layer-by-layer shifts for Mg-, N-, and O-1s (red, green and orange squares, respectively) and for Sc-2s (blue squares) core state energies for multilayers three \textbf{(a)} and six \textbf{(b)} ScN layers thick ($\Delta$E represents the value of the shift in eV). Yellow indicates the interfaces: O-Sc (IFA) and N-Mg (IFB). The zero is set at IFA. An electric field is formed from IFB to IFA producing a potential build up across the multilayer of 2-3 eV. In the ScN side of the multilayer, there are N (green) and Sc (blue) planes with formal charges 3- and 3+. In the MgO side, Mg (red) and O (orange) layers have formal charges 2+ and 2-.}\label{fig2}
\end{figure}

\section{Computational Methods}

Our electronic structure calculations were performed within density functional
theory \cite{dft,dft_2} using the all-electron, full potential code {\sc wien2k} \cite{wien} based on the augmented plane wave plus local orbitals (APW+lo) basis set.\cite{sjo}

All calculations were well converged with respect to all the parameters used. In particular, we used R$_{MT}$K$_{max}$= 7.0 (the product of the smallest of the atomic sphere radius R$_{MT}$ and the plane wave cutoff parameter K$_{max}$), which determines the size of the basis set. The chosen R$_{MT}$ are 1.95 a.u. for Sc, 1.73 a.u. for N, 1.79 a.u. for Mg and 1.69 a.u. for O. The calculations used a 43$\times$43$\times$5 $k$-mesh for the integrations over the Brillouin zone. For the structural relaxations we have used the Wu-Cohen (WC) version of the generalized gradient approximation (GGA)\cite{wu_cohen} that gives better lattice parameters for MgO than the Perdew-Burke-Ernzerhof version (PBE).\cite{gga,tran_wc} The optimized lattice parameters derived within GGA-WC for MgO and ScN are 4.23 and 4.50 \AA, respectively.

 An accurate \textit{ab initio} description requires a method able to reproduce the band gap of the individual components as well as their band alignment at the interface, both of which are generally beyond the capabilities of GGA or local density approximation (LDA). To be able to reproduce the bulk gap of ScN, we have used the semilocal potential developed by Tran and Blaha based on a modification of the Becke-Johnson potential (TB-mBJ).\cite{mbj_1,mbj_2} This is a local approximation (local in the density and the kinetic energy density) to an atomic exact-exchange potential and a screening term + GGA correlation that allows the calculation of band gaps with an accuracy similar to the much more expensive GW or hybrid methods.\cite{mbj_1,mbj_2} The TB-mBJ functional is a potential-only functional, i.e., there is no corresponding TB-mBJ exchange-correlation energy functional. In this respect it is applied as a self-energy correction, and one based on improved intra-atomic exchange processes. The band structure for bulk ScN obtained within TB-mBJ reproduces the experimental band gap and is shown in the Appendix.

Based on the agreement with the experiments for ScN bulk, TB-mBJ was the scheme applied to the  ScN/MgO (111) multilayers. We have modeled MgO/ScN(111) multilayers which are 2- to 7-ScN layers thick. A barrier of 3-4 MgO layers (about 1.5 nm thick) between ScN blocks has been used for all the calculations, checking that it is sufficient to guarantee the lack of interaction between ScN blocks. The multilayers are modeled with the in-plane lattice parameters constrained to those of MgO (fixed to the value 4.23 \AA, obtained by optimizing the cell volume within WC-GGA).  

For all the multilayers we performed calculations with fully relaxed atomic positions also optimizing the value of the $c$-lattice parameter (off-plane), i.e. allowing atomic displacements along the $c$-axis and thus relaxing the inter-plane distances for the structures with different number of ScN layers. The optimized values of the $c$-lattice parameter result in a slight increase from the value obtained by constructing the multilayers from the MgO bulk unit cell, of about 1\% to  2\% for multilayers two to seven ScN layers thick. This is expected, as the in-plane lattice parameter inside the ScN blocks is constrained to the smaller value of MgO.
The average distances between layers along the \textit{c} axis are shown in Fig. \ref{fig1} for a multilayer 5 ScN-layers thick. The main feature in the calculated relaxed geometry consisting in one shorter and one longer Sc/Mg-N/O bond length along \textit{c} can be observed.

\section{Analysis of Results}

The underlying structure of the multilayers is shown in Fig. \ref{fig1}. When growing periodically arranged superlattices two polar interfaces arise:  the so-called interface A (IFA, $n$-type) (2$^-$/3$^+$) and interface B (IFB, $p$-type) (3$^-$/2$^+$). 
If the system remains insulating (which is not obvious {\it a priori} since ScN is a small gap semiconductor)
the polar discontinuity will produce a large electrostatic potential offset between the two interfaces, 
resulting in  potential gradient reflecting the internal electric field. This effect can be tracked most
directly by following the layer-by-layer core level eigenvalues. These eigenvalues are shown in Fig. \ref{fig2}  
for multilayers with different ScN thickness. For thin ScN layers a potential difference of 2 to 3 eV on 
each block of the multilayer develops. This gradient corresponds to a local electric field of 1.1$\times$10$^7$ V/cm 
acting across the interface region, on the order of that obtained in LAO/STO 
heterostructures.\cite{eom_lao_sto,pentcheva_hole_electron} The layer-by-layer shifts in the core levels are 
about 1 eV in the thinner multilayer and are reduced when the interfaces metalize. Superlattice periodicity 
forces the potential to return to zero after it has ramped up the ScN slab.

As a consequence of this potential gradient (that scales with thickness) there is an insulator-to-metal transition as the number of ScN layers is increased. For multilayers 3 ScN layers thick the gap (1.5 eV) is widened by confinement with respect to the value in the bulk (see this band structure in the Appendix), for 4 ScN layers the gap is reduced to 0.34 eV, and from a critical thickness of 5 ScN layers the band structures are metallic. Insulator-to-metal transitions with thickness have been observed in a number of oxide-based heterostructures in the literature, both polar and non-polar, including open-shell and closed-shell materials, such as multilayers of VO$_2$/TiO$_2$,\cite{vpardo_mit_tio2vo2} SrVO$_3$/SrTiO$_3$,\cite{sto_svo_exp, sto_svo_dft_vpardo,sto_svo_dim_crossover} LaVO$_3$/SrTiO$_3$, \cite{assman} and the above mentioned LaAlO$_3$/SrTiO$_3$.\cite{mannhart, eom_lao_sto}

\subsection{Electronic structure} 
\begin{figure}
\includegraphics[width=0.9\columnwidth,draft=false]{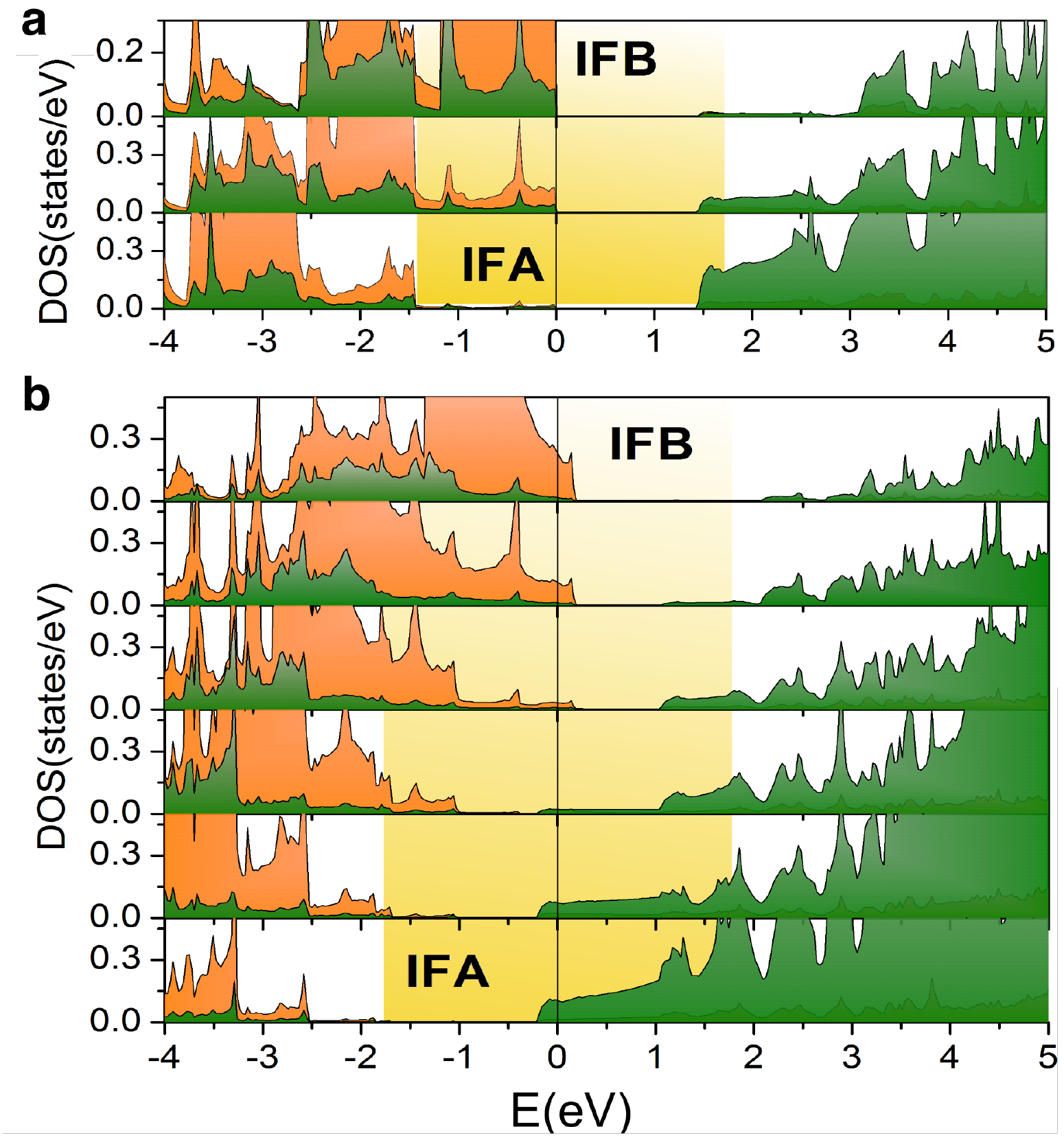}
\caption{(Color online) Layer resolved density of states for Sc/N atoms across the ScN slab from IFB to IFA for ScN/MgO(111) multilayers three \textbf{(a)} and six \textbf{(b)} ScN-layers-thick.  N-$p$ states in orange, Sc-$d$ states in green. The insulator-to-metal transition with thickness and the shift in the N-$p$ and Sc-$d$ states (of almost 1 eV/layer) as moving from IFA to IFB can be observed. The yellow patch is shaded in brightness in proportion to the gap.}\label{fig3}
\end{figure}

The Sc and N projected density of states (DOS) for three and six ScN layers are shown in Fig. \ref{fig3} where the the insulator-to-metal transition with thickness is evident. The nearly rigid upward shift consistent with the above described potential gradient can be seen by focusing on the N-$2p$ bands (orange) that move towards the Fermi level from IFA to IFB. 
The insulating state obtained for thinner ScN layers is shown in Fig. \ref{fig3}(a). The gap is formed between occupied N-$p$ states and unoccupied Sc-$d$ states.

For multilayers with a thicker ScN block both interfaces host metallic states as shown
in Fig. \ref{fig3}(b). N-$p$ and Sc-$d$ bands overlap in energy at IFB and IFA, respectively, leading to spatially separated hole and electron conducting states. The band structure and Fermi surface plots (shown in the Appendix) indicate that the nearly circular hole pocket is centered at $\Gamma$ whereas the nearly circular electron pocket is centered at the hexagonal zone face $M$ point. This band overlap provides a multilayer realization of the polar discontinuity scenario of electronic reconstruction, originally found for LAO overlayers on an STO substrate.\cite{pentcheva_reconst, naga_reconst}  Upon metallization electrons are transferred from one interface of the unit
cell to the other, thus reducing the internal field although a substantial gradient persists
(of the order of 10$^7$ V/cm). The result is two bilayer conductors each
confined to two atomic layers (IF and IF-1), constituting a periodic array of alternating electron and hole 2DGs. The N-derived 2DHG should promote robust two carrier 2D conduction instead of insulating $p$-type interfaces.

Using a transition metal nitride provides yet other opportunities. The superconductivity observed in many samples of STO/LAO may be simply the state observed in $n$-doped STO (confined somewhat to reflect a 2D superconducting behavior), or it might be a new phenomenon due to an interfacial pairing process. Ti, Zr, V, and Nb mononitrides with a rocksalt structure were among the first discovered superconductors with $T_c$'s up to 16 K.\cite{matthias, sc_review_modern_phys} Also, electron-doped transition metal chloronitrides MNCl (M = Hf, Zr, Ti) are superconducting in the 15-26 K range,\cite{MNCl} putting them among the top classes of high temperature superconductors. The observation of superconductivity at a transition metal nitride interface (as in this ScN/MgO system) is an exciting possibility, and if discovered would support that superconductivity observed at STO/LAO interfaces is derivative of bulk STO superconductivity. 

Excitons (electron-hole bound states) are a topic of great importance in condensed matter physics and play a central role in solar energy conversion.\cite{millis,el_hole_cond, acs_excitons,assman}
For four and fewer ScN layers a small gap remains, providing the platform for formation of excitons that are indirect
in $k$-space and whose components are separated in real space, giving the possibility of long-lived excitons.
Bristowe \textit{et al}. \cite{lao_sto_exciton} noted the possibility of exciton formation in the context of LaAlO$_3$/SrTiO$_3$ multilayers with alternating $p$ and $n$ interfaces, stymied because hole conduction is very difficult to achieve. 
In LAO/STO (001) overlayers a parallel electron-hole conduction has been reported only when an STO capping layer is added to protect the uppermost 2DEG.\cite{pentcheva_hole_electron, electron_hole_2d} The hole conduction relies on a surface state whereas the 2DEG is in the interfacial layer. The reported hole mobility is low and it is expected that the holes can become localized or eliminated in uncapped STO/LAO systems more strongly.
The route proposed here for the formation of proximal electron and hole gases should provide the desired platform as  compared to the standard LAO/STO heterostructure. 

In addition to the more robust hole conduction, here the system is much simpler (1/3 fewer chemical species need be introduced in the standard chambers used for thin film epitaxial growth) and that should be an advantage in terms of growth. Importantly, the problem with oxygen vacancies in STO (due to the Ti valence flexibility) is absent in MgO.

\subsection{Bilayer transport} 


To allow for additional experimental tests of the origin and nature of the conduction mechanism in these multilayers, we have calculated the temperature dependence of the thermopower $S(T)$. The expression for (diagonal) $S(T)$ within Bloch-Boltzmann transport theory\cite{Allen1988} is

\begin{equation}
\begin{split}
S_{\alpha\alpha}(T) = -\frac{k_B}{e}\frac{\int {d\varepsilon \left[\frac{\varepsilon-\mu}{k_BT}\right]
       \sigma_{\alpha\alpha}(\varepsilon})\left[-\frac{df(\varepsilon)}{d\varepsilon}\right]}{ {\int d\varepsilon ~ \sigma_{\alpha\alpha}(\varepsilon})\left[-\frac{df(\varepsilon)}{d\varepsilon}\right]}
        \end{split}
\end{equation}

\begin{eqnarray}
\sigma_{\alpha\alpha}(\varepsilon) = e^2N(\varepsilon) v_{\alpha}^2(\varepsilon) \tau(\varepsilon),
\end{eqnarray}

in terms of the energy-dependent conductivity $\sigma(\varepsilon)$, where $k_B$ is the Boltzmann constant, $e$ is the fundamental charge, $N$ the density of states, $v$ the group velocity, $f$ the distribution function, and $\tau$ the scattering time. The energy dependence of $\tau$ depends on the scattering mechanism and is usually treated as unimportant on a fine energy scale, as we do here. For low temperature, the magnitude and sign of S$_{\alpha\alpha}$ depend only on the energy derivative of $\sigma_{\alpha\alpha}$ at $\mu$ where $\mu$ is the chemical potential.

\begin{figure}
\includegraphics[width=\columnwidth,draft=false]{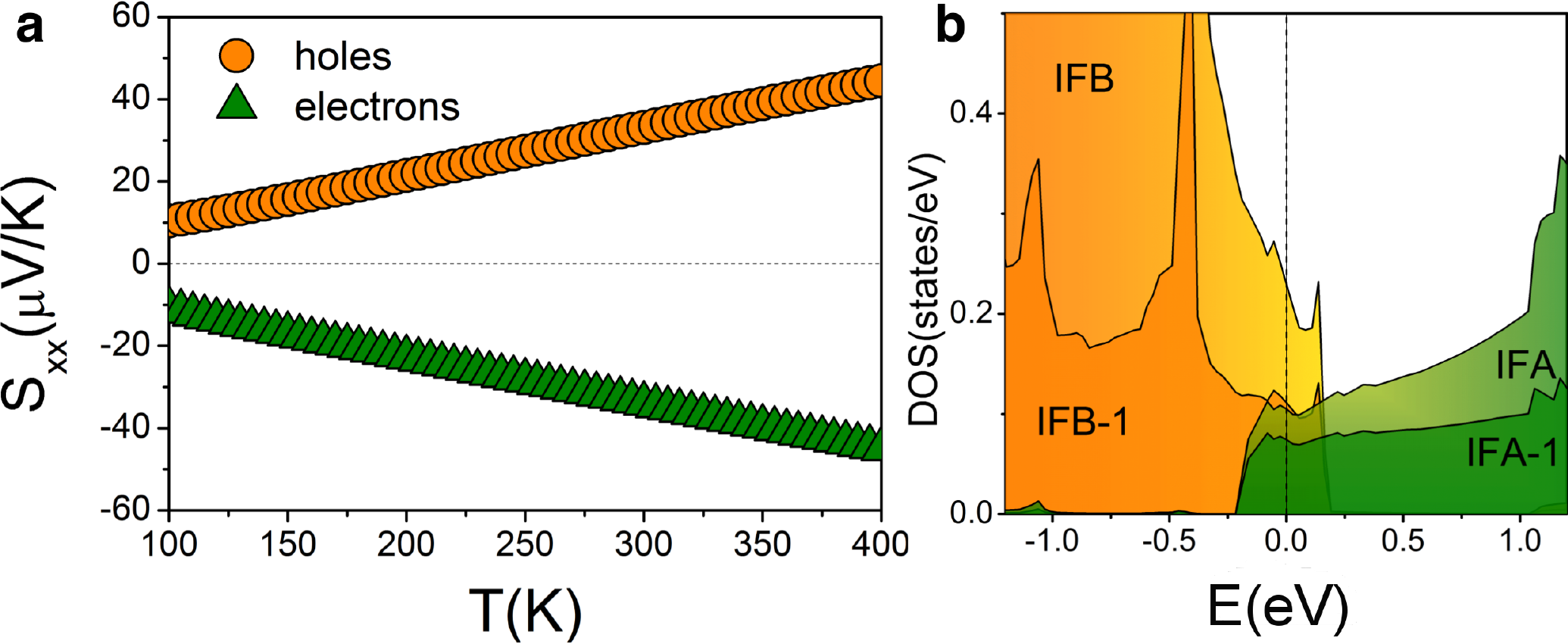}
\caption{(Color online) Bilayer conduction. (a) Temperature dependence of the in-plane thermopower for MgO/ScN multilayers six ScN layers thick for electrons and holes. The contributions from electron and hole-like bands to S$_{xx}$(T) compensate. The calculations were done by summing over only the valence
Fermi surfaces for the hole properties (orange), only over the conduction
Fermi surfaces for the electron properties (green). (b) Density of states near the Fermi level: hole-like N-$p$ states at IFB and IFB-1 (orange) and electron-like Sc-$d$ at IFA and IFA-1 (green).}\label{fig4}
\end{figure}

One should note that the calculation of the desired thermopower coefficient requires
modification of existing methods. The
band structure shown in Fig. \ref{figs2} is like that of a
compensated semimetal, with a few electron carriers and compensating hole carriers
 and Fermi energies (occupied bandwidths) of 100-200 meV.
In a conventional semimetal the carriers permeate the entire cell and make
compensating contributions to the net value. However, in the current structure
the 2D gases are distinct and it is their individual thermopowers that are of
interest. The contributions must be calculated separately, as we have done, using the Bloch-Boltzmann expressions \cite{Allen1988}, which involve only near Fermi surface quantities versus those from throughout the occupied bands 
(such as the carrier density) and takes into account the band nonparabolicity
that occurs in this system. The results
are presented in Fig. \ref{fig4}. Values of the thermoelectric
power for electrons and holes are linear with T and virtually identical in size
($|dS/dT|$ = 0.11$\mu$V/K$^2$). 
Fig. \ref{fig4}(b) shows the density of states (DOS) of electron and
hole bands separately. The differences in $N(\varepsilon)$ near $E = \varepsilon-\mu$ = 0
are evident, with the hole DOS in particular having a non-2D like behavior.


The thermopower obtained in this ScN/MgO(111) multilayer is different than some behaviors discussed
elsewhere.\cite{seebeck_sto_lao} 2DEGs can be obtained in related interfaces but involving only one type of carrier.\cite{ohta} Quantum confinement, which narrows (some) bands, may produce a thermopower enhancement.\cite{dresselhaus,mahan_sofo} Interestingly, ScN itself has been found to yield an anomalously large thermopower albeit at high doping level,\cite{thermoel_1, thermoel_init, thermoelectric_3} suggested to result from localized im-
purity states close to the Fermi level.\cite{thermoel_theory} The thermopowers of the individual 2DGs of this heterostructure represent an advance beyond previous theoretical or experimental
studies separating the 2DEG and 2DHG contributions.

Doped semiconductors, which is what each interface 2DG is, typically
display large thermopowers, and though still hotly pursued, large thermopowers
are not always what an application requires. 
This result of nearly exactly canceling electron and hole thermopowers
provides the opportunity of engineering small to negligible electrothermal
conversion, which may be needed in some applications, as opposed to the
more visible energy conversion between electrical energy and heat.
Currents that do not drag along net heat should have applications in
electronics, such as providing cooling to offset Joule heating.  All thermopower measurements have to be corrected by adding the absolute thermopower of the lead wires. Materials with negligible thermopower should allow the thermopower of other substances to be determined directly.\cite{tisc} 

As for the thermopower, the Hall coefficient $R^H$ of the separate 2DGs requires modification
of existing methods. Recall that the Hall coefficient is a Fermi surface property, 
obtained from first and second derivatives of the band dispersion $\varepsilon_k$ on the
Fermi surface. Only for quadratic, single band systems can $R^H$ be related to a carrier
density obtained from the integrated density of states. More specifically, if in the definition $R^H_{xyz}$ = $E_y/j_x B_z$
the current $j_x$ from {\it both} electron and hole 2DGs is used,
the result incorporates quantities from two separate subsystems, whereas we want
$R^H$ for each conducting layer separately.
The calculated Hall coefficient for electron and hole-like bands separately gives 
$R^H_h$ = 0.11 $\times 10^{-7}$ m$^3$/C, 
$R^H_e$ = -0.16 $\times 10^{-7}$ m$^3$/C, 
both T-independent above 200 K. These values correspond in a standard 
(but simplistic, see below) interpretation to an effective carrier density 
$n_{e}$ = 3$\times$$10^{13}$carriers/cm$^{2}$ and  
$n_{h}$ = 4$\times$$10^{13}$carriers/cm$^{2}$.  These ``Hall densities"
need not be equal,                                                
because the bands are not precisely parabolic as is assumed when $R^H$ is
converted to carrier density. 

Measured values of Hall density of oxide-only interfaces are similar to our
calculated values just above, so it is worthwhile to look more carefully at the
numbers. 
The values for LAO/STO(001) led to considerable discussion because 
the polar mismatch picture implies 
that for thick layers (i.e.
isolated interfaces) the transferred density should be 0.5 carrier per interface cell, 
which is almost an order of magnitude larger.\cite{caviglia, pentcheva_hole_electron}
Of course, for thin ScN layers as here as for most LAO/STO calculations, 
just beyond the polar mismatch
insulator-to-metal transition, only enough electrons transfer across
the layer, thereby counteracting the internal field enough to line up
the 2DEG Fermi level with the 2DHG Fermi level. 

The actual transferred charge density $n_{tr}$, from the integrated DOS, is
0.09 $e$ per interface cell, or 1.1$\times$10$^{14}$/cm$^2$,
transferred from the two nearly degenerate
$\Gamma$-centered hole bands to the two symmetry-related, $M$-centered electron
bands. This factor of three difference can be reconciled by noting that the
behavior of $N(E)$ at the Fermi level (Fig.~\ref{fig4}) is not typical of quadratic
bands ({\it i.e.} which would be constant) and that $R^H$ is sensitive to the
first and second derivatives of $\varepsilon_k$ on the Fermi surfaces. The numbers
also highlight a difference of this ScN/MgO(111) system compared to LAO/STO(001):
the areal density of transition metal ions is nearly three times larger,
with neighbors in the rocksalt (111) plane being $a/\sqrt{2}$ and close-packed,
versus in perovskite (001) the separation is $a$ and not close-packed. The
lattice constants do not differ greatly.  Thus rocksalt (111) interfaces
offer the possibility of much higher densities as well as 2DHGs that actually
conduct.

\section{summary}

We have proposed a novel approach to achieve two dimensional, two carrier bilayer conducting systems by 
moving away from oxygen for hole conduction. Designing by choice of anion and by orientation
of the interface, and studying with first principles calculations, 
hybrid oxide-nitride multilayers are proposed as a platform for the realization of 
parallel two dimensional, two carrier (electron+hole) gases. The basic concept builds on 
a slab of a narrow gap transition metal nitride (B) sandwiched between a wide gap insulating 
oxide (A). The polar discontinuities at the A and B interfaces provide an internal electric 
field that raises the N-$p$ valence band maximum above the Sc-$d$ conduction band minimum,
with resulting charge transfer metalizing both interfaces.  The use of the narrow gap nitride 
ScN means that both signs of carriers live in ScN. Electrically MgO is inert, shifting the active component
from an oxide to a nitride.

\acknowledgments

V. P. thanks MINECO for project MAT2013-44673-R, the Xunta de Galicia through project EM2013/037, and Spanish Government for financial support through the Ram\'on y Cajal
Program. W.E.P. acknowledges many conversations with R. Pentcheva on the theory and
phenomenology of oxide interfaces and multilayers. 
A.S.B. 
 and  W.E.P were supported by Department of Energy Grant No. DE-FG02-04ER46111. 

\section{appendix}
\subsection{Electronic structure of bulk ScN}

\begin{figure}[H]
\center
\includegraphics[width=\columnwidth,draft=false]{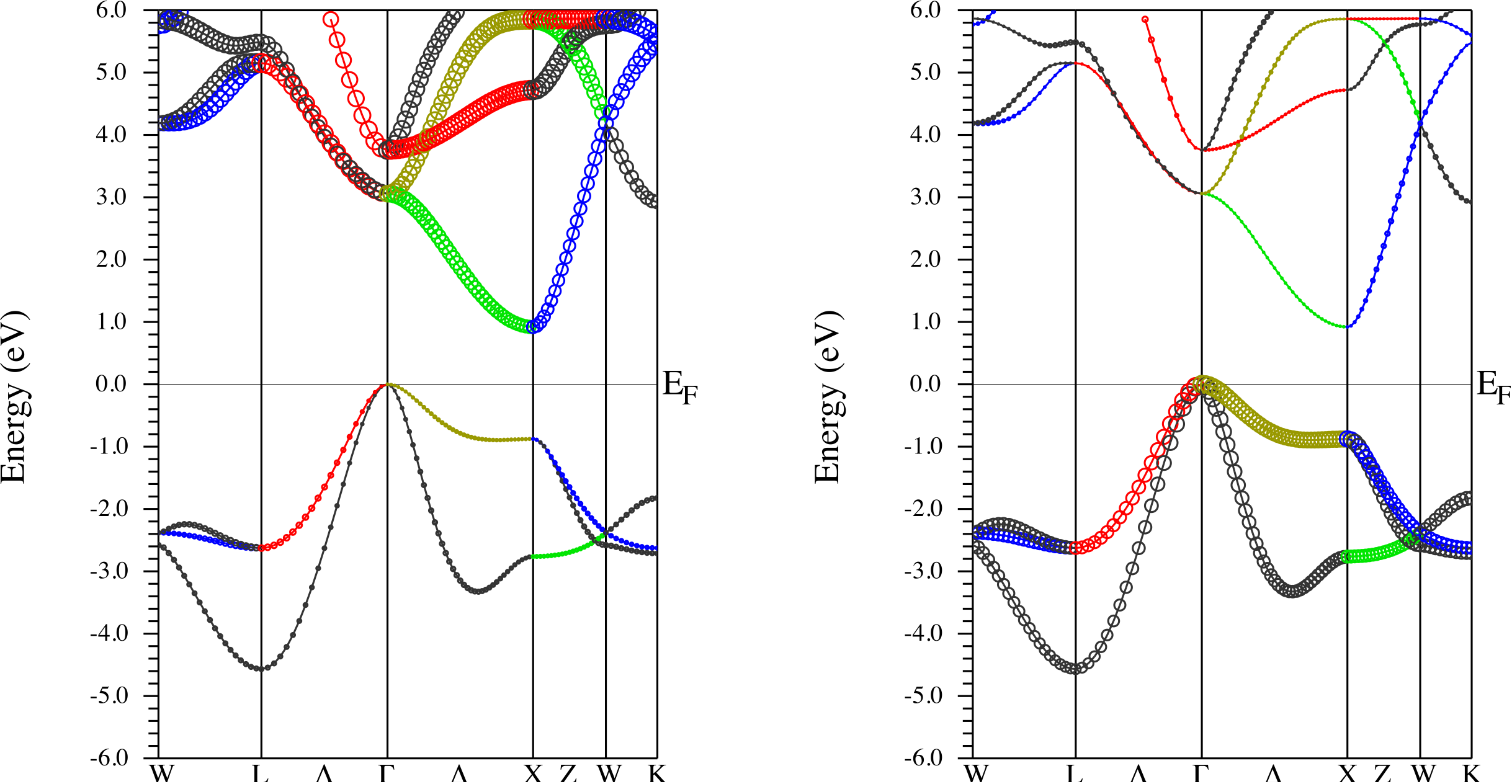}
\caption{{(Color online) Electronic structure of bulk ScN.} Band structure with orbital character 
shown by circle size, for ScN using the TB-mBJ potential. Left panel: Sc $3d$; right panel: N $2p$.  
The colors help to identify corresponding bands across the two panels. }\label{bulk_bs}
\end{figure}

ScN is a nonmagnetic semiconducting transition metal mononitride with a band gap of 0.9 eV that has remarkable physical properties such as high chemical stability, high melting point (2600$\degree$C), and large mechanical hardness (H= 21 GPa).\cite{gall_2} 

A number of first principles calculations have been performed on ScN. Early density functional theory-based calculations employing the LDA \cite{semimetal_es_3} or X$\alpha$\cite{semimetal_es, semimetal_es_2} approximations predicted ScN to be a semimetal. Exact exchange and screened exchange calculations confirm that ScN is a semiconductor with an indirect $\Gamma$-$X$ band gap, consistent with experiments. However, the calculated band gap of 1.6 eV is significantly larger than the experimental value of 0.9 eV.\cite{gall_es} The LDA+U\cite{sic} scheme improves over GGA or LDA in the study of systems containing correlated electrons by introducing the on-site Coulomb repulsion $U$ applied to localized electrons (in this case Sc-$3d$ states). For ScN, LDA/GGA+$U$ calculations using a $U$ value as high as 6 eV are only able to open an indirect gap of 0.40 to 0.55 eV, respectively, far from the one determined experimentally.\cite{dft_u}
 LDA calculations complemented with estimated quasiparticle corrections and calculations of the optical response give an indirect band gap of 0.9 eV with a first direct gap at $X$ of 2 eV.\cite{lambrecht}
 Quasiparticle G$_0$W$_0$ calculations predict ScN to have an indirect band gap between $\Gamma$-$X$ of 0.99 eV.\cite{gw_2} Also hybrid functionals show the opening of a band gap of 0.9 eV.\cite{alling}
 
Figure \ref{bulk_bs} shows the band structure with band character plot (N and Sc character highlighted) for ScN obtained within TB-mBJ. The band gap is formed between occupied N-$2p$ and unoccupied Sc-$3d$ states and the experimental value of the gap is reproduced: an indirect band gap of 1.0 eV between $\Gamma$-$X$ is obtained as well as the direct gap of 2.1 eV at $X$. Thus TB-mBJ provides a computationally efficient method of producing and accurate band structure, and it has been applied for the results discussed in this manuscript.

\subsection{Electronic structure of MgO/ScN(111) multilayers}

\begin{figure}
\center
\includegraphics[width=\columnwidth,draft=false]{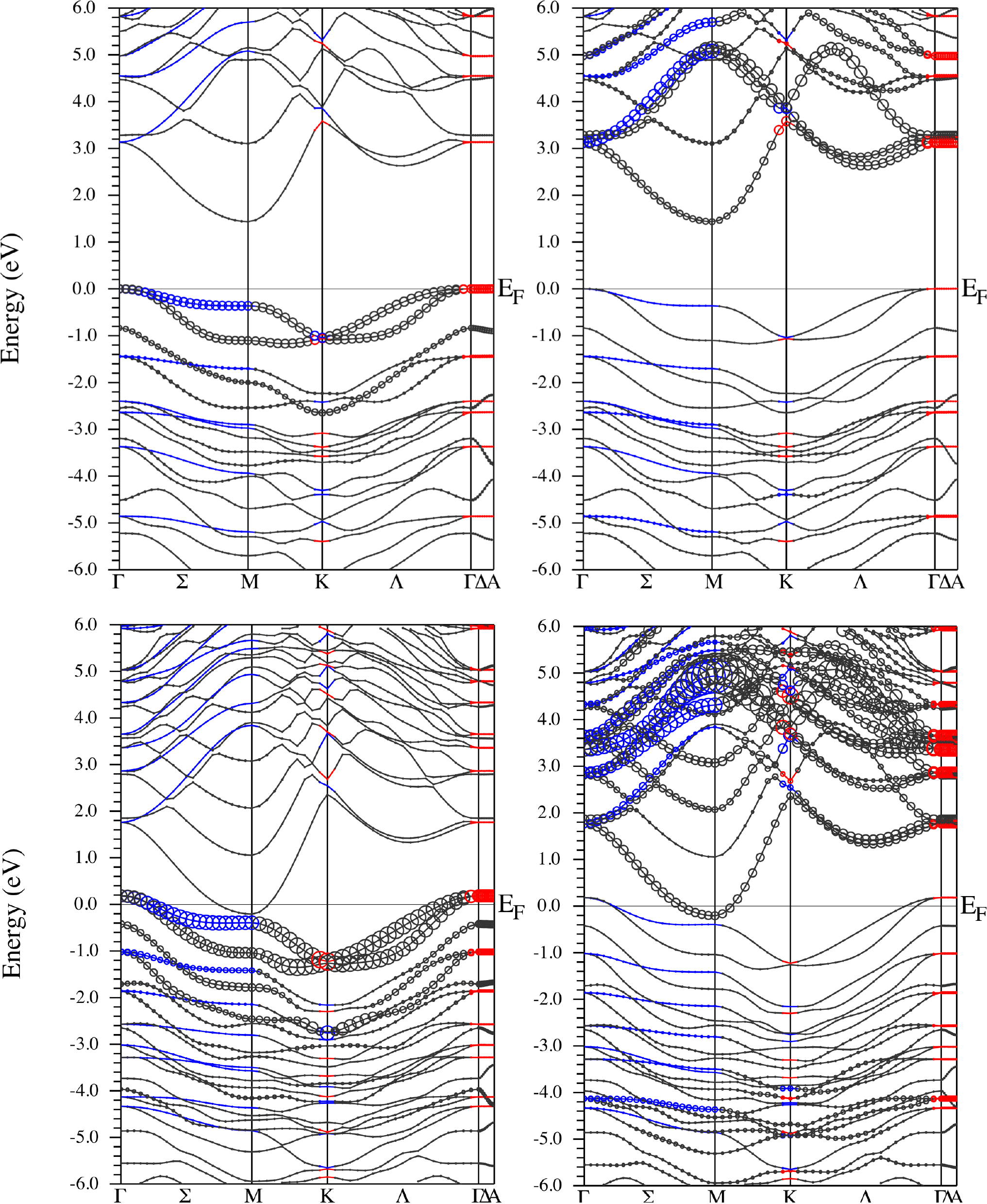}\\
\caption{(Color online) Band structure plot with band character emphasized for MgO/ScN (111) multilayers obtained within GGA-TB-mBJ. Sc-$d$ atom at IFA(right) and N-$p$ levels at IFB (left) for ScN/MgO(111) multilayers three (top panel), where a band gap of 1.5 eV persists, and six (bottom panel) ScN-layers-thick, where a band overlap of 0.4 eV can be seen.The colors indicate the irreducible representation of the eigenvalues. Whenever two bands cross they have different colors denoting the different irreducible representations, the connectivity of a band across a band crossing can be seen just following the color of the band.}\label{figs2}
\end{figure}

Figure \ref{figs2} shows the band structures with band character plot that complement Fig. 3 in the main text,
which shows the corresponding DOS. The bands crossing the Fermi level are indeed N-$p$ at  IFB and Sc-$d$ at the IFA. 

\begin{figure}
\center
\includegraphics[width=\columnwidth,draft=false]{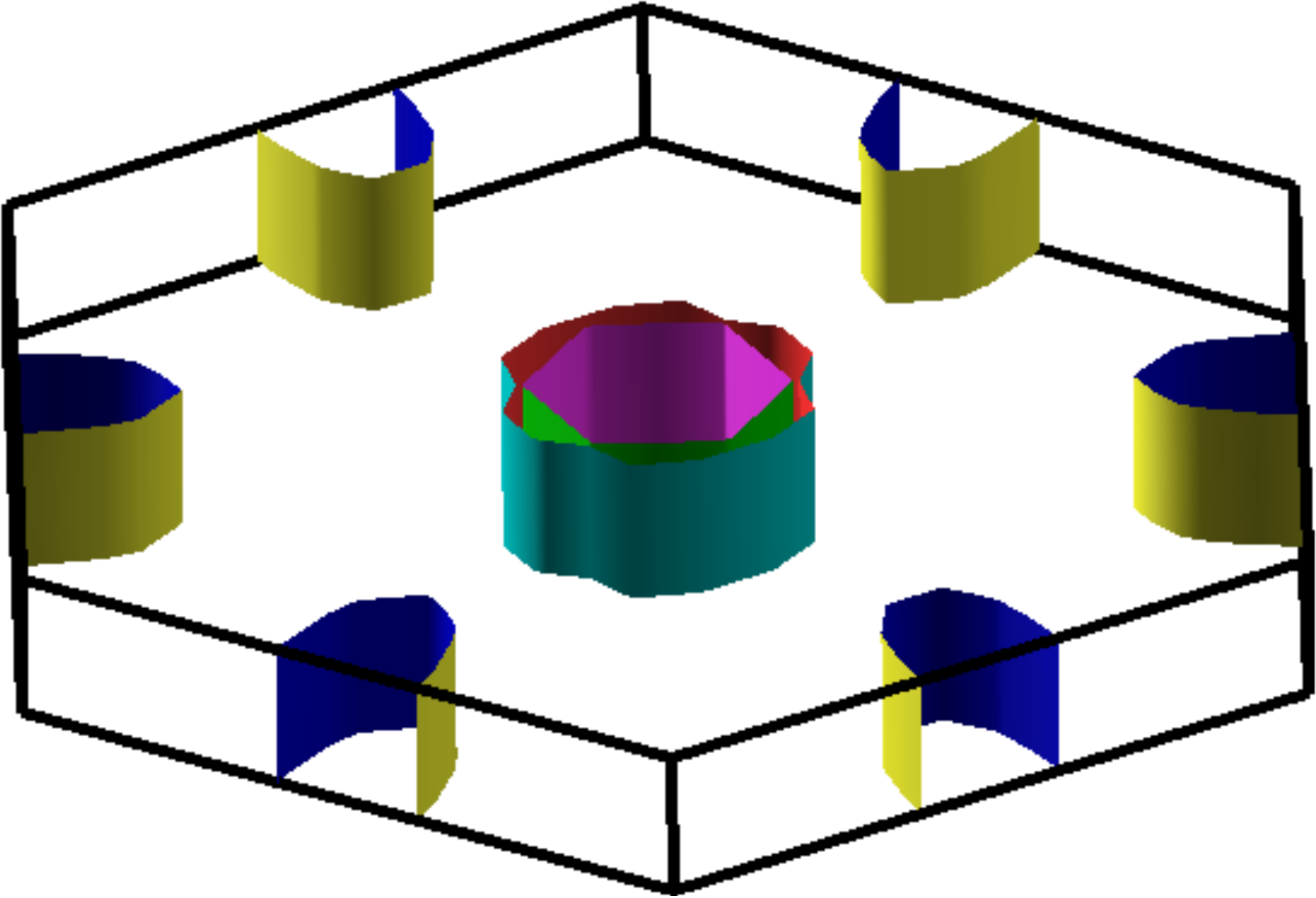}
\caption{(Color online){Fermi surface plots for hole and electron pockets obtained for ScN/MgO(111) multilayers 6 ScN layers thick.}  The concentric, nearly circular hole pockets are centered at $\Gamma$ whereas the nearly circular electron pocket is centered at the hexagonal zone face $M$ point.}\label{figs3}
\end{figure}

Figure \ref{figs3} shows that these bands give rise to nearly circular Fermi surfaces coming from the electron (centered at $M$) and hole pockets (centered at $\Gamma$) respectively, as discussed in the main text.

\end{document}